\documentclass[12pt]{article}

\usepackage{amsmath,amssymb,array,calc,rotating,epsfig,psfrag, amscd, datetime, comment, color, bm, mathrsfs}

\usepackage{color}
\usepackage[
      colorlinks=true,
      urlcolor=blue,    
      filecolor=blue,     
      citecolor=red,
      pdfstartview=FitV,
       bookmarksopen=true    
      ]{hyperref}
\usepackage[left=2.5cm,top=2.5cm,right=2.5cm,nohead]{geometry}
\numberwithin{equation}{section}

\newcommand{\nc}{\newcommand}

\definecolor{cardinal}{rgb}{0.6,0,0}
\definecolor{darkgreen}{rgb}{0,0.5,0}
\definecolor{golden}{rgb}{0.92, 0.7, 0}
\definecolor{midnight}{rgb}{0, 0, 0.5}
\definecolor{darkblue}{rgb}{0.2, 0, 0.8}
\definecolor{myblue}{rgb}{0.17, 0.13, .71}

\definecolor{slategrey}{rgb}{.44,.50,.56}
\definecolor{lightslategrey}{rgb}{.47,.53,.6}
\definecolor{silver}{rgb}{.75,.75,.75}
\nc{\SlateGrey}{\color{slategrey}}
\nc{\LightSlateGrey}{\color{lightslategrey}}
\nc{\Silver}{\color{silver}}

\definecolor{thistle}{RGB}{216,191,216}
\nc{\Thistle}{\color{thistle}}
\definecolor{blueviolet}{RGB}{138,43,226}
\nc{\BlueViolet}{\color{thistle}}
\definecolor{mediumpurple}{RGB}{147,112,219}
\nc{\MediumPurple}{\color{mediumpurple}}

\nc{\ra}{\rightarrow} 
\nc{\lra}{\leftrightarrow} 
\nc{\Ra}{\Rightarrow} 
\nc{\LRa}{\Leftightarrow} 
\nc{\blp}{{\big (}}
\nc{\brp}{{\big )}}
\nc{\Blp}{{\Big (}}
\nc{\Brp}{{\Big )}}
\nc{\bglp}{{\bigg (}}
\nc{\bgrp}{{\bigg )}}
\nc{\Bglp}{{\Bigg (}}
\nc{\Bgrp}{{\Bigg )}}
\nc{\slb}{{\rm [}}
\nc{\srb}{{\rm ]}}
\nc{\bslb}{{\rm \big [}}
\nc{\bsrb}{{\rm \big ]}}
\nc{\Bslb}{{\rm \Big [}}
\nc{\Bsrb}{{\rm \Big ]}}

\def\al{\alpha}

\def\eps{\epsilon}
\nc{\veps}{\varepsilon}
\def\gam{\gamma}

\def\lam{\lambda}
\def\om{\omega}

\nc{\vphi}{\varphi}
\def\tha{\theta}

\def\sig{\sigma}

\def\Gam{\Gamma}

\def\Lam{\Lambda}
\def\Om{\Omega}
\def\Sig{\Sigma}

\def\coeff#1#2{\relax{\textstyle {#1 \over #2}}\displaystyle}

\nc{\myvspace}{\rule[-1em]{0pt}{2.5em}}
\nc{\bea}{\begin{eqnarray}}
\nc{\eea}{\end{eqnarray}}
\nc{\be}{\begin{equation}}
\nc{\ee}{\end{equation}}
\nc{\barr}{\begin{array}}
\nc{\earr}{\end{array}}

\nc{\co}{{\cal o}}

\nc{\cA}{{\cal A}}
\nc{\cB}{{ \cal B}}
\def\cC{{\cal C}}
\def\cD{{\cal D}}

\nc{\cF}{{\cal F}}
\nc{\cG}{{\cal G}}
\def\cH{{\cal H}}
\def\cI{{\cal I}}

\def\cK{{\cal K}}
\nc{\cL}{{\cal L}}
\nc{\cM}{{\cal M}}

\def\cN{{\cal N}}

\def\cO{{\cal O}}

\nc{\cQ}{{\cal Q}}
\nc{\cR}{{\cal R}}
\def\cS{{\cal S}}

\def\cV{{\cal V}}
\def\cV{{\cal V}}
\def\cW{{\cal W}}

\def\cZ{{\cal Z}}
\nc{\cQd}{\cQ^{\dagger}}
\nc{\cRd}{\cR^{\dagger}}
\nc{\BB}{{\mathbb B}}
\nc{\CC}{{\mathbb C}}
\nc{\DD}{{\mathbb D}}
\nc{\EE}{{\mathbb E}}
\nc{\FF}{{\mathbb F}}
\nc{\GG}{{\mathbb G}}
\nc{\HH}{{\mathbb H}}
\nc{\JJ}{{\mathbb J}}
\nc{\MM}{{\mathbb M}}
\nc{\NN}{{\mathbb N}}
\nc{\PP}{{\mathbb P}}
\nc{\QQ}{{\mathbb Q}}
\nc{\RR}{{\mathbb R}}
\nc{\UU}{{\mathbb U}}
\nc{\ZZ}{{\mathbb Z}}
\nc{\calone}{{\mathbb 1}}

\nc{\half}{\frac{1}{2}}
\nc{\quarter}{\coeff{1}{4}}
\nc{\del}{\partial}

\nc{\delbar}{\bar\partial}
\nc{\thalf}{\frac{t}{2}}
\nc{\Spin}{\operatorname{Spin}}
\nc{\SO}{\operatorname{SO}}

\nc{\Sp}{{\rm Sp}}
\nc{\com}[2]{{ \left[ #1, #2 \right] }}
\nc{\acom}[2]{{ \left\{ #1, #2 \right\} }}
\nc{\rr}{\rightarrow}
\nc{\p}{\partial}
\nc{\LT}{{\LL_\T}}
\nc{\Tr}{{\rm Tr}}
\nc{\tr}{{\rm tr}}
\nc{\Adag}{A^{\dagger}}
\nc{\AdagI}{A^{\dagger I}}
\nc{\AdagJ}{A^{\dagger J}}
\nc{\AdagK}{A^{\dagger K}}
\nc{\AdagL}{A^{\dagger L}}
\nc{\AdagM}{A^{\dagger M}}
\nc{\Bdag}{B^{\dagger}}
\nc{\BdagI}{B^{\dagger}_I}
\nc{\BdagJ}{B^{\dagger}_J}
\nc{\BdagK}{B^{\dagger}_K}
\nc{\BdagL}{B^{\dagger}_L}
\nc{\BdagM}{B^{\dagger}_M}
\nc{\Cdag}{C^{\dagger}}
\nc{\CdagI}{C^{\dagger I}}
\nc{\CdagJ}{C^{\dagger J}}
\nc{\CdagK}{C^{\dagger K}}
\nc{\Ddag}{D^{\dagger}}
\nc{\DdagI}{D^{\dagger I}}
\nc{\DdagJ}{D^{\dagger J}}
\nc{\DdagK}{D^{\dagger K}}
\nc{\bva}{\breve{a}}
\nc{\bvb}{\breve{b}}
\nc{\bvc}{\breve{c}}
\nc{\bvd}{\breve{d}}
\nc{\bve}{\breve{e}}
\nc{\bvf}{\breve{f}}
\nc{\bvg}{\breve{g}}
\nc{\bvh}{\breve{h}}
\nc{\bvi}{\breve{i}}
\nc{\bvj}{\breve{j}}
\nc{\bvk}{\breve{k}}
\nc{\bvl}{\breve{l}}
\nc{\bvm}{\breve{m}}
\nc{\bvn}{\breve{n}}
\nc{\bvo}{\breve{o}}
\nc{\bvp}{\breve{p}}
\nc{\brvq}{\breve{q}}
\nc{\bvr}{\breve{r}}
\nc{\bvs}{\breve{s}}
\nc{\bvt}{\breve{t}}
\nc{\bvu}{\breve{u}}
\nc{\bvv}{\breve{v}}
\nc{\bvw}{\breve{w}}
\nc{\bvx}{\breve{x}}
\nc{\bvy}{\breve{y}}
\nc{\bvz}{\breve{z}}

\nc{\bvA}{\breve{A}}
\nc{\bvB}{\breve{B}}
\nc{\bvC}{\breve{C}}
\nc{\bvD}{\breve{D}}
\nc{\bvE}{\breve{E}}
\nc{\bvF}{\breve{F}}
\nc{\bvG}{\breve{G}}
\nc{\bvH}{\breve{H}}
\nc{\bvI}{\breve{I}}
\nc{\bvJ}{\breve{J}}
\nc{\bvK}{\breve{K}}
\nc{\bvL}{\breve{L}}
\nc{\bvM}{\breve{M}}
\nc{\bvN}{\breve{N}}
\nc{\bvO}{\breve{O}}
\nc{\bvP}{\breve{P}}
\nc{\bvQ}{\breve{Q}}
\nc{\bvR}{\breve{R}}
\nc{\bvS}{\breve{S}}
\nc{\bvT}{\breve{T}}
\nc{\bvU}{\breve{U}}
\nc{\bvV}{\breve{V}}
\nc{\bvcV}{\breve{\cV}}
\nc{\bvW}{\breve{W}}
\nc{\bvX}{\breve{X}}
\nc{\bvY}{\breve{Y}}
\nc{\bvZ}{\breve{Z}}

\nc{\ul}[1]{{\underline{#1}}}

\nc{\tal}{\widetilde{\alpha}}
\nc{\tbeta}{\widetilde{\beta}}
\nc{\ttha}{\tilde{\theta}}
\nc{\ttau}{\tilde{\tau}}
\nc{\tTha}{\tilde{\Theta}}
\nc{\tphi}{\tilde{\phi}}
\nc{\tsig}{\tilde{\sig}}
\nc{\tom}{\widetilde{\om}}
\nc{\tOm}{\widetilde{\Om}}
\nc{\tlam}{\widetilde{\lam}}
\nc{\tLam}{\tilde{\Lam}}
\nc{\tSig}{\widetilde{\Sig}}
\nc{\tPhi}{\tilde{\Phi}}
\nc{\tPhibar}{\ol{\tPhi}}
\nc{\tPi}{\widetilde{\Pi}}
\nc{\tpsi}{\widetilde{\psi}}
\nc{\tPsi}{\tilde{\Psi}}
\nc{\tgam}{\widetilde{\gam}}
\nc{\tGam}{\widetilde{\Gam}}
\nc{\tzeta}{\tilde{\zeta}}
\nc{\tZeta}{\tilde{\Zeta}}
\nc{\teta}{\widetilde{\eta}}
\nc{\teps}{\tilde{\eps}}
\nc{\tveps}{\tilde{\veps}}
\nc{\tEta}{\tilde{\Eta}}
\nc{\tchi}{\tilde{\chi}}
\nc{\tChi}{\tilde{\Chi}}
\nc{\txi}{\tilde{\xi}}
\nc{\tXi}{\widetilde{\Xi}}
\nc{\tnu}{\tilde{\nu}}
\nc{\tmu}{\tilde{\mu}}

\nc{\ta}{\tilde a}
\nc{\tb}{\tilde b}
\nc{\tc}{\tilde c}
\nc{\te}{\tilde e}
\nc{\tf}{\widetilde f}
\nc{\tg}{\widetilde g}
\nc{\ti}{\tilde i}
\nc{\tj}{\tilde j}
\nc{\tk}{\widetilde k}
\nc{\tl}{\tilde l}
\nc{\tm}{\widetilde m}
\nc{\tn}{\tilde n}
\nc{\tp}{\tilde{p}}
\nc{\tq}{\widetilde{q}}
\nc{\trr}{{\tilde r}}
\nc{\ts}{{\tilde s}}
\nc{\tu}{{\tilde u}}
\nc{\tv}{{\tilde v}}
\nc{\tw}{{\tilde w}}
\nc{\tx}{{\tilde x}}
\nc{\ty}{{\tilde y}}
\nc{\tz}{\tilde z}
\nc{\tA}{{\widetilde A}}
\nc{\tAbar}{{\ol \tA}}
\nc{\tB}{{\widetilde B}}
\nc{\tC}{{\widetilde C}}
\nc{\tD}{{\widetilde D}}
\nc{\tE}{{\widetilde E}}
\nc{\tF}{{\widetilde F}}
\nc{\tG}{{\widetilde G}}
\nc{\tcG}{{\widetilde \cG}}
\nc{\tH}{{\widetilde H}}
\nc{\tcH}{{\widetilde \cH}}
\nc{\tI}{{\widetilde I}}
\nc{\tcI}{{\widetilde \cI}}
\nc{\tJ}{{\widetilde J}}
\nc{\tJbar}{{\ol {\tilde J}}}
\nc{\tK}{{\widetilde K}}
\nc{\tL}{{\widetilde L}}
\nc{\tcL}{{\widetilde \cL}}
\nc{\tcLbar}{{\ol \tcL}}
\nc{\tM}{{\widetilde M}}
\nc{\tN}{{\widetilde N}}
\nc{\tcN}{{\widetilde \cN}}
\nc{\tP}{{\widetilde P}}
\nc{\tQ}{{\widetilde Q}}
\nc{\tR}{{\widetilde R}}
\nc{\tcR}{{\widetilde \cR}}
\nc{\tS}{\widetilde{S}}
\nc{\tcS}{\widetilde{\cS}}
\nc{\tT}{\widetilde{T}}
\nc{\tU}{\widetilde{U}}
\nc{\tUU}{\widetilde{\UU}}
\nc{\tV}{\widetilde{V}}
\nc{\tcVbar}{\ol{\widetilde{\cV}}}
\nc{\tW}{\widetilde{W}}
\nc{\tcF}{\widetilde{{\cal F}}}
\nc{\tX}{\widetilde{X}}
\nc{\tY}{\widetilde{Y}}
\nc{\tcZ}{\tilde{\cZ}}
\nc{\tcZbar}{\ol{\tcZ}}

\nc{\ha}{\hat a}
\nc{\hb}{\hat b}
\nc{\hc}{\widehat c}
\nc{\hd}{\widehat d}
\nc{\he}{\widehat e}
\nc{\hf}{\widehat f}
\nc{\hg}{\widehat g}
\nc{\hh}{\widehat h}
\nc{\hn}{\widehat n}
\nc{\hp}{\widehat p}
\nc{\hq}{\widehat q}
\nc{\hr}{\widehat r}
\nc{\hs}{\widehat s}
\nc{\hu}{\widehat u}
\nc{\hv}{\widehat v}
\nc{\hw}{\widehat w}
\nc{\bhw}{{\bf \hw}}
\nc{\hx}{\widehat x}
\nc{\hy}{\widehat y}
\nc{\hz}{\widehat z}
\nc{\zhat}{\hat z}
\nc{\hA}{\widehat{A}}
\nc{\hB}{\widehat{B}}
\nc{\hC}{\widehat{C}}
\nc{\hD}{\widehat{D}}
\nc{\hcD}{\widehat{\cD}}
\nc{\hE}{\widehat{E}}
\nc{\hF}{\widehat{F}}
\nc{\hcF}{\widehat{\cF}}
\nc{\hG}{\widehat{G}}
\nc{\hcG}{\widehat{\cG}}
\nc{\hH}{\widehat{H}}
\nc{\hI}{\widehat{I}}
\nc{\hcI}{\widehat{\cI}}
\nc{\hJ}{\widehat{J}}
\nc{\hK}{\widehat{K}}
\nc{\hcK}{\widehat{\cK}}
\nc{\hL}{\widehat{L}}
\nc{\hcL}{\widehat{\cL}}
\nc{\hM}{\widehat M}
\nc{\hcM}{\widehat{\cM}}
\nc{\hN}{\widehat{N}}
\nc{\hO}{\widehat{O}}
\nc{\hcO}{\widehat{\cO}}
\nc{\hP}{\widehat{P}}
\nc{\hQ}{\widehat{Q}}
\nc{\hcQ}{\widehat{\cQ}}
\nc{\hcR}{\widehat{\cR}}
\nc{\hR}{\widehat{R}}
\nc{\hS}{\widehat{S}}
\nc{\hcS}{\widehat{\cS}}
\nc{\hT}{\widehat{T}}
\nc{\hU}{\widehat{U}}
\nc{\hV}{\widehat V}
\nc{\hcV}{\widehat \cV}
\nc{\tcV}{\widetilde{\cV}}
\nc{\hX}{\widehat X}
\nc{\hcZ}{\widehat \cZ}
\nc{\hcZbar}{\ol{\widehat \cZ}}

\nc{\heta}{\widehat{\eta}}
\nc{\hal}{\widehat \alpha}
\nc{\hbeta}{\widehat \beta}
\nc{\heps}{\widehat \eps}
\nc{\hphi}{\widehat{\phi}}
\nc{\hkap}{\hat{\kappa}}
\nc{\hchi}{\widehat{\chi}}
\nc{\hpsi}{\widehat{\psi}}
\nc{\hgam}{\widehat{\gam}}
\nc{\hPhi}{\hat{\Phi}}
\nc{\hPsi}{\hat{\Psi}}
\nc{\hGam}{\hat{\Gam}}
\nc{\omhat}{\widehat{\om}}
\nc{\Omhat}{\widehat{\Om}}
\nc{\hsig}{\widehat{\sig}}
\nc{\hSig}{\widehat{\Sig}}
\nc{\htha}{\hat{\tha}}
\nc{\hrho}{\widehat{\rho}}
\nc{\hdel}{\widehat{\del}}
\nc{\hdelbar}{\ol{\hdel}}
\nc{\hnabla}{\widehat{\nabla}}

\nc{\w}{\wedge}

\nc{\vb}{\vec b}
\nc{\vc}{\vec c}
\nc{\vd}{\vec d}
\nc{\ve}{\vec e}
\nc{\vf}{\vec f}
\nc{\vg}{\vec g}
\nc{\vh}{\vec h}
\nc{\vk}{\vec k}
\nc{\vl}{\vec l}
\nc{\vm}{\vec m}
\nc{\vn}{\vec n}
\nc{\vp}{\vec p}
\nc{\vq}{\vec q}
\nc{\vr}{\vec r}
\nc{\vs}{\vec s}
\nc{\vv}{\vec v}
\nc{\vw}{\vec w}
\nc{\vx}{\vec x}
\nc{\vy}{\vec y}
\nc{\vz}{\vec z}

\nc{\vB}{\vec B}
\nc{\vC}{\vec C}
\nc{\vD}{\vec D}
\nc{\vE}{\vec E}
\nc{\vF}{\vec F}
\nc{\vG}{\vec G}
\nc{\vH}{\vec H}
\nc{\vP}{\vec P}
\nc{\vQ}{\vec Q}
\nc{\vR}{\vec R}
\nc{\vS}{\vec S}
\nc{\vV}{\vec V}
\nc{\vW}{\vec W}
\nc{\vX}{\vec X}
\nc{\vY}{\vec Y}
\nc{\vZ}{\vec Z}

\nc{\val}{\vec \al}
\nc{\vbeta}{\vec \beta}
\nc{\vmu}{\vec \mu}
\nc{\vtha}{\vec \theta}
\nc{\vecphi}{\vec \phi}
\nc{\vecvphi}{\vec \vphi}
\nc{\vecsig}{\vec \sig}

\nc{\ol}{\overline}
\nc{\abar}{\ol{a}}
\nc{\bbar}{\ol{b}}
\nc{\cbar}{\ol{c}}
\nc{\dbar}{\ol{d}}
\nc{\ebar}{\ol{e}}
\nc{\fbar}{\ol{f}}
\nc{\gbar}{\ol{g}}
\nc{\ibar}{\ol{\imath}}
\nc{\jbar}{\ol{\jmath}}
\nc{\kbar}{\ol{k}}
\nc{\lbar}{\ol{l}}
\nc{\mbar}{\ol{m}}
\nc{\nbar}{\ol{n}}
\nc{\pbar}{\ol{p}}
\nc{\qbar}{\ol{q}}
\nc{\rbar}{\ol{r}}
\nc{\sbar}{\ol{s}}
\nc{\ubar}{\ol{u}}
\nc{\vbar}{\ol{v}}
\nc{\wbar}{\ol{w}}
\nc{\xbar}{\ol{x}}
\nc{\ybar}{\ol{y}}
\nc{\zbar}{\ol{z}}

\nc{\Abar}{\ol{A}}
\nc{\cAbar}{\ol{\cA}}
\nc{\Bbar}{\ol{B}}
\nc{\cBbar}{\ol{\cB}}
\nc{\Cbar}{\ol{C}}
\nc{\cCbar}{\ol{\cC}}
\nc{\Dbar}{\ol{D}}
\nc{\Ebar}{\ol{E}}
\nc{\hEbar}{\ol{\hE}}
\nc{\Fbar}{\ol{F}}
\nc{\Gbar}{\ol{G}}
\nc{\Jbar}{\ol{J}}
\nc{\Kbar}{\ol{K}}
\nc{\cKbar}{\ol{\cK}}
\nc{\Lbar}{\ol{L}}
\nc{\cLbar}{\ol{\cL}}
\nc{\Mbar}{\ol{M}}
\nc{\Nbar}{\ol{N}}
\nc{\Pbar}{\ol{P}}
\nc{\Qbar}{\ol{Q}}
\nc{\Rbar}{\ol{R}}
\nc{\Sbar}{\ol{S}}
\nc{\Tbar}{\ol{T}}
\nc{\Ubar}{\ol{U}}
\nc{\Vbar}{\ol{V}}
\nc{\cVbar}{\ol{\cV}}
\nc{\Wbar}{\ol{W}}
\nc{\cWbar}{\ol{\cW}}
\nc{\Xbar}{{\overline X}}
\nc{\Ybar}{{\overline Y}}
\nc{\Zbar}{{\overline Z}}
\nc{\cZbar}{{\overline \cZ}}

\nc{\epsbar}{\ol{\epsilon}}
\nc{\albar}{\ol{\al}}
\nc{\Albar}{\ol{\Al}}
\nc{\betabar}{\ol{\beta}}
\nc{\Betabar}{\ol{\Beta}}
\nc{\deltabar}{\ol{\delta}}
\nc{\etabar}{\ol{\eta}}
\nc{\lambar}{\ol{\lambda}}
\nc{\kapbar}{\ol{\kappa}}
\nc{\zetabar}{\ol{\zeta}}
\nc{\Zetabar}{\ol{\Zeta}}
\nc{\taubar}{\ol{\tau}}
\nc{\Taubar}{\ol{\Tau}}
\nc{\psibar}{\ol{\psi}}
\nc{\Psibar}{\ol{\Psi}}
\nc{\tpsibar}{\ol{\tpsi}}
\nc{\tPsibar}{\ol{\tPsi}}
\nc{\phibar}{\ol{\phi}}
\nc{\Phibar}{\ol{\Phi}}
\nc{\chibar}{\ol{\chi}}
\nc{\sigbar}{\ol{\sig}}
\nc{\Sigbar}{\ol{\Sig}}
\nc{\mubar}{\ol{\mu}}
\nc{\nubar}{\ol{\nu}}
\nc{\rhobar}{\ol{\rho}}
\nc{\ombar}{\ol{\om}}
\nc{\Ombar}{\ol{\Om}}
\nc{\Deltabar}{\ol{\Delta}}
\nc{\Thetabar}{\ol{\Theta}}
\nc{\xibar}{\ol{\xi}}
\nc{\Xibar}{\ol{\Xi}}

\nc{\Dthbar}{\ol{\rm D3}}

\nc{\fdot}{\dot{f}}
\nc{\gdot}{\dot{g}}
\nc{\pdot}{\dot{p}}
\nc{\qdot}{\dot{q}}
\nc{\rdot}{\dot{r}}
\nc{\sdot}{\dot{s}}
\nc{\tdot}{\dot{t}}
\nc{\udot}{\dot{u}}
\nc{\vdot}{\dot{v}}
\nc{\wdot}{\dot{w}}
\nc{\xdot}{\dot{x}}
\nc{\xddot}{\ddot{x}}
\nc{\ydot}{\dot{y}}
\nc{\zdot}{\dot{z}}
\nc{\yddot}{\ddot{y}}

\nc{\Adot}{\dot{A}}
\nc{\Bdot}{\dot{B}}
\nc{\Cdot}{\dot{C}}
\nc{\dotD}{\dot{D}}
\nc{\Fdot}{\dot{F}}
\nc{\Pdot}{\dot{P}}
\nc{\Qdot}{\dot{Q}}
\nc{\Rdot}{\dot{R}}
\nc{\Sdot}{\dot{S}}
\nc{\Tdot}{\dot{T}}
\nc{\Udot}{\dot{U}}
\nc{\Vdot}{\dot{V}}
\nc{\Wdot}{\dot{W}}

\nc{\taudot}{\dot{\tau}}
\nc{\phidot}{\dot{\phi}}
\nc{\psidot}{\dot{\psi}}
\nc{\chidot}{\dot{\chi}}
\nc{\Gamdot}{\dot{\Gam}}
\nc{\sinp}{s_{\phi}}
\nc{\cosp}{c_{\phi}}
\nc{\tanp}{t_{\phi}}
\nc{\spone}{s_{\phi_1}}
\nc{\cpone}{c_{\phi_1}}
\nc{\tpone}{t_{\phi_1}}
\nc{\sptwo}{s_{\phi_2}}
\nc{\cptwo}{c_{\phi_2}}
\nc{\tptwo}{t_{\phi_2}}
\nc{\spth}{s_{\phi_3}}
\nc{\cpth}{c_{\phi_3}}
\nc{\tpth}{t_{\phi_3}}
\nc{\calp}{c_{\al}}
\nc{\salp}{s_{\al}}

\nc{\csch}{{\rm csch}}
\nc{\sech}{{\rm sech}}

\nc{\cothzlami}{\coth(z-\lam_i)}
\nc{\coshzlami}{\cosh(z-\lam_i)}
\nc{\sinhzlami}{\sinh(z-\lam_i)}

\nc{\cothzlamj}{\coth(z-\lam_j)}
\nc{\coshzlamj}{\cosh(z-\lam_j)}
\nc{\sinhzlamj}{\sinh(z-\lam_j)}

\nc{\cothlamij}{\coth(\lam_i-\lam_j)}
\nc{\coshlamij}{\cosh(\lam_i-\lam_j)}
\nc{\sinhlamij}{\sinh(\lam_i-\lam_j)}

\nc{\bah}{{\mathbf {\hat{A}}}}
\nc{\bX}{{\mathbf X}}
\nc{\ba}{{\bf a}}
\nc{\bb}{{\bf b}}
\nc{\bc}{{\bf c}}
\nc{\bd}{{\bf d}}
\nc{\bg}{{\bf g}}
\nc{\bk}{{\bf k}}
\nc{\bl}{{\bf l}}
\nc{\bn}{{\bf n}}
\nc{\bo}{{\bf o}}
\nc{\bp}{{\bf p}}
\nc{\bq}{{\bf q}}
\nc{\br}{{\bf r}}
\nc{\bs}{{\bf s}}
\nc{\bt}{{\bf t}}
\nc{\bu}{{\bf u}}
\nc{\bv}{{\bf v}}
\nc{\bw}{{\bf w}}
\nc{\bx}{{\bf x}}
\nc{\by}{{\bf y}}
\nc{\bz}{{\bf z}}

\nc{\bH}{{\bf H}}
\nc{\bP}{{\bf P}}
\nc{\bQ}{{\bf Q}}

\nc{\bom}{{\bf \om}}
\nc{\bombar}{{\mathbf \ombar}}
\nc{\bPhi}{{\bf \Phi}}
\nc{\bSig}{{\bf \Sig}}

\nc{\rma}{{\rm a}}
\nc{\rmb}{{\rm b}}
\nc{\rmc}{{\rm c}}
\nc{\rmd}{{\rm d}}
\nc{\rmg}{{\rm g}}
\nc{\rk}{{\rm k}}
\nc{\rml}{{\rm l}}
\nc{\rmm}{{\rm m}}
\nc{\rmn}{{\rm n}}
\nc{\rmo}{{\rm o}}
\nc{\rmp}{{\rm p}}
\nc{\rmq}{{\rm q}}
\nc{\rmr}{{\rm r}}
\nc{\rms}{{\rm s}}
\nc{\rmt}{{\rm t}}
\nc{\rmu}{{\rm u}}
\nc{\rmv}{{\rm v}}
\nc{\rmw}{{\rm w}}
\nc{\rmx}{{\rm x}}
\nc{\rmy}{{\rm y}}
\nc{\rmz}{{\rm z}}

\nc{\dal}{\dot{\al}}
\nc{\thadot}{\dot{\tha}}
\nc{\thab}{\bar{\theta}}
\nc{\thal}{\theta^{\al}}
\nc{\thdal}{\bar{\theta}^{\dal}}

\nc{\thsigthm}{\tha \sigma^m \thab}
\nc{\thsigthn}{\tha \sigma^n \thab}

\nc{\Dal}{D_{\al}}
\nc{\Ddal}{\bar{D}_{\dal}}
\nc{\CDal}{{\cal D}_{\al}}
\nc{\CDdal}{\bar{\cal D}_{\dal}}

\nc{\eq}[1]{{(\ref{#1})}}
\nc{\eqtwo}[2]{{(\ref{#1},\ref{#2})}}
\nc{\eqthree}[3]{(\ref{#1},\ref{#2},\ref{#3})}
\nc{\eqfour}[4]{(\ref{#1},\ref{#2},\ref{#3},\ref{#4})}
\nc{\eqfive}[5]{(\ref{#1},\ref{#2},\ref{#3},\ref{#4,\ref{#5}})}
\nc{\non}{\nonumber}
\nc{\Fzero}{F_{(0)}}
\nc{\Ftwo}{F_{(2)}}
\nc{\Ffour}{F_{(4)}}
\nc{\Fone}{F_{(1)}}
\nc{\Fthree}{F_{(3)}}
\nc{\Ffive}{F_{(5)}}
\nc{\Fn}{F_{(n)}}
\nc{\Fp}{F_{(p)}}

\nc{\tFzero}{\tF_{(0)}}
\nc{\tFtwo}{\tF_{(2)}}
\nc{\tFfour}{\tF_{(4)}}
\nc{\tFone}{\tF_{(1)}}
\nc{\tFthree}{\tF_{(3)}}
\nc{\tFfive}{\tF_{(5)}}
\nc{\tFn}{\tF_{(n)}}
\nc{\tFp}{\tF_{(p)}}

\nc{\Czero}{C_{(0)}}
\nc{\Ctwo}{C_{(2)}}
\nc{\Cfour}{C_{(4)}}
\nc{\Cone}{C_{(1)}}
\nc{\Cthree}{C_{(3)}}
\nc{\Cfive}{C_{(5)}}
\nc{\Cn}{C_{(n)}}

\nc{\equ}{{\rm eq}}
\def\Im{{\rm Im \hspace{0.5mm} }}

\def\Re{{\rm Re \hspace{0.5mm}}}

\nc{\vol}{{\rm vol}}
\nc{\Ainf}{A_{\infty}}
\nc{\End}{{\rm End}}
\nc{\Ext}{{\rm Ext}}
\nc{\IIB}{{\rm IIB}}
\nc{\Ad}{{\rm Ad}}
\nc{\IIA}{{\rm IIA}}
\nc{\AdS}{{\rm AdS}}
\nc{\CFT}{{\rm CFT}}
\nc{\diag}{{\rm diag}}
\nc{\Log}{{\rm Log}}
\nc{\Mat}{{\rm Mat}}
\nc{\mat}{{\rm mat}}
\nc{\cRic}{{\cR ic}}
\nc{\hcRic}{\widehat{\cRic}}
\nc{\Dir}{{\rm Dir}}

\nc{\Dslash}{\ensuremath \raisebox{0.025cm}{\slash}\hspace{-0.32cm} D}
\nc{\cDslash}{\ensuremath \raisebox{0.025cm}{\slash}\hspace{-0.32cm} \cD}
\nc{\omslash}{\om\!\!\!/}
\nc{\delslash}{\del\!\!\! /}
\nc{\Hslash}{H\!\!\!\! /}
\nc{\Kslash}{K\!\!\!\!/}
\nc{\Yslash}{Y\!\!\!\!/}

\nc{\no}{\!:\!\!}
\nc{\ointdz}{\oint\frac{dz}{2\pi i}}
\nc{\ointdzone}{\oint\frac{dz_1}{2\pi i}}
\nc{\ointdztwo}{\oint\frac{dz_2}{2\pi i}}
\nc{\ointdzb}{\oint\frac{d\zbar}{2\pi i}}
\nc{\ointdzbone}{\oint\frac{d\zbar_1}{2\pi i}}
\nc{\ointdzbtwo}{\oint\frac{d\zbar_2}{2\pi i}}
\nc{\dz}{\frac{dz}{2\pi i}}
\nc{\dzb}{\frac{d\zbar}{2\pi i}}
\nc{\bpm}{\begin{pmatrix}}
\nc{\epm}{\end{pmatrix}}
 \nc{\bitem}{\begin{itemize}}
 \nc{\eitem}{\end{itemize}}
 \nc{\exercise}{\vskip 2mm \noindent {\bf Exercise:}}
 
\begin{document}

\vspace{0.5cm}
\begin{center}
\baselineskip=13pt {\LARGE \bf{On the On-Shell: \\ The Action of
AdS$_4$ Black Holes}\\}
 \vskip1.5cm 
Nick Halmagyi and  Shailesh Lal\\ 
 \vskip0.5cm
\textit{Laboratoire de Physique Th\'eorique et Hautes Energies,\\
Universit\'e Pierre et Marie Curie, CNRS UMR 7589, \\
F-75252 Paris Cedex 05, France}\\
\vskip0.5cm

\end{center}

\begin{abstract}
We compute the on-shell action of static, BPS  black holes in AdS$_4$ from $\cN=2$ gauged supergravity coupled to vector multiplets and show that it is equal to minus the entropy of the black hole. Holographic renormalization is used to demonstrate that with appropriate boundary conditions on the scalar fields, the divergent and finite contributions from the asymptotic boundary vanish. The entropy arises from the extrinsic curvature on $\Sig_g\times S^1$ evaluated at the horizon, where $\Sig_g$ may have any genus $g\geq 0$. This provides a clarification of the equivalence between the partition function of the twisted ABJM theory on $\Sig_g\times S^1$ and the entropy of the dual black hole solutions. It also demonstrates that the complete entropy resides on the AdS$_2\times \Sig_g$ horizon geometry, implying the absence of hair for these gravity solutions.
\end{abstract}

\section{Introduction}

Holography has provided a robust framework for explaining the Bekenstein-Hawking entropy of a black hole from microscopic considerations. For black holes in asymptotically flat space-times, the essential breakthrough was \cite{Strominger:1996sh} but the subsequent development of AdS/CFT \cite{Maldacena:1997re} has lead to a far deeper understanding of this line of research. The starting point of such work is typically supersymmetric black holes and in AdS$_4$, static-BPS black holes with spherical horizon were only found somewhat recently \cite{Cacciatori:2009iz}. In \cite{Benini:2015eyy, Benini:2016rke} the entropy of these black holes\footnote{as well as the the dyonic black holes of \cite{Halmagyi:2013qoa, Halmagyi:2014qza}} has been reproduced from the holographically dual field theory using fundamentally different methods than those employed in \cite{Strominger:1996sh}; namely by localization and without recourse to the Cardy formula of a two-dimensional CFT.

The purpose of the current work is to further clarify the relationship between the macroscopic computations of the entropy and the microscopic calculations of \cite{Benini:2015eyy, Benini:2016rke}. The holographic dictionary relates the boundary partition function to the on-shell action of the gravitational theory, not the entropy. In this work we consider supersymmetric (and extremal) dyonically charged black holes in $\cN=2$ FI gauged supergravity coupled to $n_v$ vector multiplets and show that the on-shell action coincides with the entropy. Our methods are quite general and do not require the explicit black hole solution beyond the leading order AdS$_4$ configuration. For the ReissnerÐ-Nordstr\"om AdS$_4$ black hole (with a single magnetic charge, constant scalar fields and a negatively curved horizon of genus greater than one) an equivalent result was obtained \cite{Azzurli:2017kxo} by explicit calculation using the BPS limit of the finite temperature solution. The black holes we study here have comparatively non-trivial profiles for the metric and scalar fields (as well as possibly spherical horizons), making explicit evaluation of the on-shell action somewhat unintuitive even if it were possible. We employ a more formal strategy to derive this equivalence and do not need to reference the explicit form of the BPS or finite temperature solution. We anticipate this strategy will generalize to a variety of related settings.

Crucial to this calculation is holographic renormalization\footnote{see the interesting work \cite{Gnecchi:2014cqa} for  work on holographic renormalization of finite temperature black holes in AdS$_4$.} \cite{Bianchi:2001kw}. The first step of holographic renormalization is to cancel the divergences and for supersymmetric theories with scalar fields there are few results in the literature, although recently some interesting works have appeared \cite{Freedman:2013oja, Freedman:2016yue}. The divergent boundary counterterms we utilize are similar in form to those presented in \cite{Papadimitriou:2007sj}: a certain superpotential term cancels the cubic divergences and a term formed from the boundary Ricci scalar coupled to the scalar fields cancels the linear divergences. The finite counterterms are typically a more vexing issue but here we find that the basic principles espoused in \cite{Klebanov:1999tb} work simply and effectively: for BPS solutions the Legendre transform of the scalars fields enforces Neumann boundary counditions and cancels all finite contributions from asymptotic infinity. 

Having renormalized the on-shell action there remains a contribution to the on-shell action from the horizon. This comes from the extrinsic curvature on $\Sig_g\times S^1$ evaluated at the horizon\footnote{The Riemann surface $\Sig_g$ has genus $g\geq 0$.} and we show this to be precisely equal to the entropy of the black hole. For BPS black holes this is the only contribution to the on-shell action and we have
\be
\tcS_{\rm on-shell}{\Big |}_{\rm BPS}=-\frac{A}{4G_{N}}\,,
\ee
where $\tcS_{\rm on-shell}$ is the Legendre transform of $\cS_{\rm on-shell}$ defined in \eq{Legendre}.

\section{Supersymmetric Black Holes in AdS$_4$}
In this section we review some facts about quarter-BPS black holes in AdS$_4$ with boundary $\Sig_g\times S^1$. The reader who is already familiar with this literature may wish to skip to section \ref{sec:holofree}.

\subsection{$\cN=2$ FI-gauged supergravity}

The bulk action of four dimensional $\cN=2$ FI-gauged supergravity with $n_v$ vector mulitplets is\footnote{See appendix \ref{app:Special} for detail of the notation of the special geometry quantities}
\bea
\cS_{\rm bulk}=\frac{1}{8\pi G_N}\int\,d^4x\sqrt{g} \left( \frac{1}{2} R- g_{i\jbar} \p_{\mu} z^i \p^{\mu} \bar z^{\jbar} +  \frac{1}{4} \cI_{\Lambda\Sigma}F^{\Lambda}_{\mu \nu}F^{\Sigma\,\mu \nu}+\frac{1}{4} \cR_{\Lam \Sig} \frac{\eps^{\mu\nu\rho\sig} }{2\sqrt{g}}  F_{\mu\nu}^\Lam F_{\rho\sig}^\Sig-V_g 
\right) \non \\
\label{N2Bulk}
\eea
where the scalar potential is given by
\be
V_g=|\cL_i|^2-3|\cL|^2
\ee
and to which should be added the Gibbons-Hawking-York \cite{Gibbons:1976ue, York:1972sj} boundary term 
\bea
\cS_{{\rm GHY}}&=& -\frac{1}{8\pi G_N}\int_{M_{\infty}} d^3x \sqrt{h} K \label{GHYterm}\,.
\eea
The equation of motion for the metric which follows from \eq{N2Bulk} will be utilized later so we give it explicitly here
\bea
-(R_{\mu\nu}-\frac{1}{2}g_{\mu\nu}) &=& g_{\mu\nu}  V_g + g_{\mu\nu}  g_{i\jbar} \del^\sig z^i \del_\sig \zbar^{\ibar} -2g_{i\jbar} \del_\mu z^i \del_\nu \zbar^{\jbar}  \non \\
&& - \frac{1}{4} g_{\mu\nu} \cI_{\Lam\Sig}  F_{\rho\sig}^\Lam F^{\rho\sig\, \Sig}+ \cI_{\Lam\Sig} F^\Lam_{\mu\sig} F_{\nu}^{\ \sig\,\Sig} \label{Einy1}\,.
\eea
The equations of motion for the scalar fields and Maxwell's equation will not be needed in this work.

\subsection{The black hole ansatz}
The (static) black hole ansatz in Euclidean signature is
\bea
ds^2&=& e^{2U}dt^2 + e^{-2U} dr^2 + e^{2(V-U)} d\Sig_g^2\,, \\
p^\Lam &=& \frac{1}{4\pi} \int_{\Sig_g}F^\Lam\,,\qquad q_\Lam = \frac{1}{4\pi} \int_{\Sig_g}G_\Lam \label{charges}
\eea
where the dual field strength is 
\bea
G_{\Lam}&=& \cR_{\Lam\Sig} F^\Sig  - \cI_{\Lam \Sig} *_4 F^{\Sig}\,.
\eea
The Riemann surface $\Sig_g$ is  $(S^2,T^2,\HH^2/\Gam)$  and has genus $g$, the metric $d\Sig_g^2$ is the uniform metric of curvature $\kappa=(1,0,-1)$ respectively. The scalar fields are radially dependent $z^i=z^i(r)$ and the gauge fields contribute to the solution just through the conserved charges \eq{charges}.
The charges and gauge couplings naturally form symplectic vectors $\cQ$ and $\cG$ from which there are two symplectic invariants $\cL$ and $\cZ$:
\bea
\cQ &=& \bpm P^\Lam \\ Q_{\Lam } \epm \,,\qquad 
\cG = \bpm g^\Lam \\ g_{\Lam } \epm \,,\qquad
\cL= \langle \cG, \cV \rangle\,,\qquad \cZ = \langle \cQ,\cV \rangle\,.
\eea

Einstein's equation on this ansatz reduce to\footnote{This form of the equations can be found in \cite{Toldo:2012ec} in signature $(+---)$ but we have corrected a minus sign error in the coefficient of $V_{BH}$ which is crucial for our analysis in the next section.}
\bea
V_{BH}&=&\frac{1}{2}e^{2(V-U)} \Bslb \kappa -\frac{1}{2}(e^{2V})''+\blp e^{2(V-U)} (e^{2U})' \brp' \Bsrb \label{eom1} \\
V_g&=& \frac{1}{2} e^{2(U-V)} \Bslb \kappa -\frac{1}{2} (e^{2V})'' \Bsrb \label{eom2}\\
g_{i\jbar} z^{i\prime} \bar z^{\jbar\prime} &=& - e^{U-V}(e^{V-U})'' \label{eom3}
\eea
where we use the standard definition of $V_{BH}$ 
\be
V_{BH}=|\cZ_i|^2 + |\cZ|^2=-\frac{1}{2} \cQ^T \cM \cQ\,.
\ee
We will use the explicit form of \eq{eom1}-\eq{eom3} in section \ref{sec:onshell}.

\subsection{BPS solutions}

The BPS equations for the preservation of two real supersymmetries (commonly referred to as quarter-BPS) are\footnote{Half-BPS solutions which fit in this ansatz and preserve eight real supersymmetries also exist \cite{Duff:1999gh} but are nakedly singular. A general analysis of such solutions with scalar fields has not yet been completed.} \cite{Cacciatori:2009iz, Dall'Agata:2010gj}
\bea
2 e^{2V}  \Bslb \Im\blp e^{-i\psi}e^{-U}\cV \brp \Bsrb' &=&   8 e^{2(V-U)} \Re\blp e^{-i\psi} \cL\brp  \Re \blp e^{-i\psi}\cV \brp - \cQ - e^{2(V-U)} \Om \cM \cG \label{BPS1}\\
\blp e^V \brp' &=& 2  e^{V-U} \Im \blp  e^{-i\psi} \cL\brp \label{BPS2} \\
\psi'+\cA_r&=& - 2e^{-U}\Re (e^{-i\psi} \cL)\label{SpinorEq} \\
\langle \cG , \cQ \rangle &=&  -\kappa \label{DiracQu}
\eea
where $\psi$ is the phase of the supersymmetry parameter. 
It will prove useful to reproduce an equivalent form of \eq{BPS1}
\bea
2 \bslb\Re ( e^Ue^{-i\psi}\cV)\bsrb'&=& e^{2(U-V)} \Om \cM \cQ + \cG \label{BPS3}\,.
\eea
as well as 
\be
(e^{V-U})' =  e^{V-2U} \Im(e^{-i\psi} \cL )+ e^{-V} \Re(e^{-i\psi} \cZ) \label{BPSW}
\ee 
which can be derived from \eq{BPS1} and \eq{BPS2}.

Starting from \cite{Cacciatori:2009iz}, the solution for BPS black holes in AdS$_4$ has been developed \cite{Dall'Agata:2010gj, Hristov:2010ri, Gnecchi:2013mta, Klemm:2015xda} and in \cite{Halmagyi:2014qza} a general solution for dyonically charged, AdS$_4$ black holes in FI-gauged supergravity (with general dyonic gaugings) was derived. This solution assumes that $\cM_v$ is a homogeneous space and is presented in terms of the quartic invariant\footnote{We use the same conventions as \cite{Katmadas:2014faa} where the quartic invariant was introduced in the study of these black holes.} $I_4$.  We will not use this explicit solution in much detail but note that $e^{2V}$ is a quartic polynomial. In the solution of \cite{Cacciatori:2009iz} this quartic has a pair of double roots, while for the more general solutions of \cite{Halmagyi:2014qza} it has a single double root (required for all zero-temperature solutions).

The general solution for BPS horizon configurations of the form AdS$_2\times \Sig_g$ was found in \cite{Halmagyi:2013qoa} (for any homogeneous $\cM_v$) and the entropy was shown to equal
\bea
S&=& \frac{\vol(\Sig_g)}{4G_N}\sqrt{\frac{I_4(\cG,\cG,\cQ,\cQ)\pm \sqrt{I_4(\cG,\cG,\cQ,\cQ)^2-16 I_4 (\cG)I_4(\cQ)})}{8 I_4(\cG)} }\,.
\eea
It was also found that the BPS conditions impose an additional constraint on the charges in terms of the gauge couplings:
\bea 
0&=& 4I_4(\cG) I_{4}(\cG,\cQ,\cQ,\cQ)^2+4I_{4}(\cQ) I_{4}(\cQ,\cG,\cG,\cG)^2 \non \\
&&-  I_4(\cG,\cQ,\cQ,\cQ) I_4(\cG,\cG,\cQ,\cQ)I_{4}(\cQ,\cG,\cG,\cG)   \label{constraint}\,.
\eea

The Dirac quantization condition due to the charged gravitino is
\be
\langle \cG , \cQ \rangle \in \ZZ \label{DiracQu2}
\ee 
but supersymmetry enforces something stronger \eq{DiracQu}. One can break supersymmetry explicitly but preserve extremality by satisfying \eq{DiracQu2} but not \eq{DiracQu} and indeed the solutions in the literature satisfy the equations of motion when the charges are altered to satisfy \eq{DiracQu2} instead of \eq{DiracQu} while keeping all metric modes and scalar fields unchanged, a fact which was first noted in \cite{Klemm:2012yg, Gnecchi:2012kb} Throughout this work, while we constantly refer to BPS solutions, everything will apply equally well to solutions which satisfy \eq{DiracQu2} but not \eq{DiracQu}.

\subsection{AdS$_4$ boundary conditions}\label{sec:AdS}

The essential computations in this paper will not require the explicit form of the black hole solutions beyond the asymptotic boundary conditions which we now review. The leading order solution is the supersymmetric AdS$_4$ vacuum and can be given without reference to the assumption that $\cM_v$ is homogenous:
\bea
e^{2V} &=&\frac{r^4}{R_{{\rm AdS}}^2}\,,\qquad e^{2U} =\frac{r^2}{R_{{\rm AdS}}^2}\,,\qquad 
\Re (e^{-i\psi} \cV) =\frac{R_{{\rm AdS}} }{2}\cG
\eea
where
\be
R_{{\rm AdS}}=\frac{1}{I_4(\cG)^{1/4}}\,.
\ee
When $\cM_v$ is homogeneous we can additionally infer that
\be
\Im (e^{-i\psi} \cV) = \frac{R_{{\rm AdS}}^3  }{4}  I'_4(\cG)\,. 
\ee

\subsection{STU model}\label{sec:STU}
The most well studied black holes in AdS$_4$ with non-trivial scalar fields comes from the the STU truncation \cite{Duff:1999gh, Cvetic:1999xp} of the de-Wit Nicolai $\cN=8$ gauged supergravity \cite{deWit:1982ig}. This corresponds to the data
\bea
F=-\frac{X^1X^2X^3}{X^0}\,,\quad g^\Lam=-\bpm 0 \\ g\\g\\g \epm\,,\quad g_\Lam=\bpm g \\ 0\\0\\0 \epm\,. \label{STUframe1}
\eea
This can be easily rotated to the more familiar symplectic frame
\bea
F=-2i \sqrt{X^0X^1X^2X^3}\,,\quad g^\Lam=0\,,\quad g_\Lam=\bpm g \\ g\\g\\g \epm\,.\label{STUframe2}
\eea
which is perhaps better loved by most than the frame \eq{STUframe1} due the vanishing of the magnetic gaugings.
The quartic invariant is by construction frame invariant 
\be
I_4(\cG)=4g^4\,.
\ee
The solution of \cite{Cacciatori:2009iz} is given in the frame \eq{STUframe2} where the charges are purely magnetic.

While in this work we have little need for the explicit form of the black hole solutions, which can be found in the various papers already mentioned, we now review the dimension of the solution space. There are eight charges (four magnetic and four electric) and the BPS magnetic solutions of \cite{Cacciatori:2009iz} have four non-trivial magnetic charges subject to the single constraint \eq{DiracQu}. The constraint \eq{constraint} vanishes identically for the solutions of \cite{Cacciatori:2009iz}. It was shown in \cite{Halmagyi:2013uza} that using a $U(1)^2$ symmetry (axion shifts) of the scalar potential \cite{Cvetic:2000tb} one can generate two electric charges in addition to the magnetic charges of \cite{Cacciatori:2009iz} while preserving supersymmetry and not changing the form of the metric. A third $U(1)$ symmetry was identified which generates a third electric charge but  this is equivalent to the extra (discrete) parameter obtained by enforcing \eq{DiracQu2} instead of \eq{DiracQu}. The general BPS solution was found in \cite{Halmagyi:2014qza} and this solution space is six dimensional, it satisfies \eq{DiracQu} and \eq{constraint} but one can increase the solution space to seven dimensions by enforcing \eq{DiracQu2} at the expense of \eq{DiracQu}. Finally we mention \cite{Chow:2013gba} where a solution was found for all eight charges although the subset of these solutions which preserve supersymmetry have not yet been identified. In the absence of scalar hair (which is not at all clear) it is reasonable to expect that the solutions of  \cite{Halmagyi:2014qza} are co-dimension one inside those of \cite{Chow:2013gba} but given the rather unwieldy nature of all these solutions, it seems challenging to make this precise.

\subsection{Universal black hole}

The universal black hole \cite{Romans:1991nq, Caldarelli1999} has constant scalar fields and $\kappa=-1$, it is a solution of any supergravity theory which admits a truncation to minimal gauged supergravity, thus the moniker ``universal". In the conventions of the STU model in section \ref{sec:STU} it takes the form:\footnote{In \cite{Azzurli:2017kxo} a second non-BPS parameter was considered corresponding to varying the charge. We do not see great utility for this since due to Dirac quantization, the charge cannot be varied infinitesimally.}
\bea
&& e^{2(V-U)}= r^2\,,\quad e^{2U}= \frac{1}{2}\blp 2g\, r-\frac{1}{2g\,r} \brp^2 -2\eta r\,,\\
&&R_{\rm AdS}=\frac{1}{\sqrt{2}\, g}\,,\quad p^\Lam=\frac{1}{4g}\,,\quad q_{\Lam=0}\,,
\eea
the BPS limit is $\eta=0$. For completeness we give the conventions for the scalar fields and supersymmetry parameter
\bea
&& L^\Lam= 2^{-3/2}( 1,i,i,i )^T\,, \quad M_{\Lam}=2^{-3/2}( i ,1,1,1 )^T \,,\quad z^j=i\,, \\
&& \cL = \sqrt{2}\,g \,,\quad \cZ =\frac{i}{2^{3/2}g}\,,\quad \psi=-\frac{\pi}{2}\,.
\eea
To first order in $\eta$, the horizon is at 
\be
r_h= \frac{1}{2g}+\frac{1}{\sqrt{2g}}\eta^{1/2} + \cO(\eta^{3/2})\,.
\ee

For all the BPS solutions of \cite{Halmagyi:2014qza} the metric function $e^{2V}$ is a quartic polynomial and by comparison with \cite{Chow:2013gba} it appears that the corresponding space of finite temperature solutions has 
\be
e^{2V}\ra e^{2V} -2\eta r
\ee 
for some {\it finite} paramter $\eta$, in particular $e^{2V}$ remains a quartic polynomial for finite $\eta$. Since in the BPS solutions $e^{2V}$ has a double real root, one can show that the first correction to the horizon is order $\cO(\eta^{1/2})$ and thus $\beta= \frac{1}{\eta^{1/2}} + \cO(\eta^0)$.

\section{Holographic Free Energy} \label{sec:holofree}

To compute the holographic free-energy of BPS black holes in AdS$_4$ we find it useful to compute the on-shell action in two ways: firstly by using the second order equations of motion and secondly utilizing the BPS form of the dimensionally reduced 1d action. After adding counterterms and performing a Legendre transform we will arrive at our central result \eq{mainresult}. The entropy emerges from a total derivative term identical to the Gibbons-Hawking-York boundary term but evaluated at the horizon instead of the boundary.

\subsection{On-shell action, a first look}\label{sec:onshell}

The simplified form of Einstein's equations \eq{eom1}-\eq{eom3} allow us to evaluate the integral in $\cS_{\rm bulk}$ directly. We first reduce the action to one dimension
\bea
\cS_{\rm bulk}&=& \frac{\beta \vol(\Sig_g)}{8\pi G_N} \int_{r_h}^{r_{\infty}} \Bslb  -\kappa +e^{2V} \bslb (U'-V')^2 + 2 (V')^2 + 2 V''  -U'' \bsrb   \non \\
&&+ e^{2V} g_{i\jbar } z^{i\prime} \zbar^{\jbar\prime}  + e^{2(U-V)} V_{BH} + e^{2(V-U)} V_g\Bsrb
\eea
then after some algebra we find\footnote{The domain of Euclidean time is $t\in (0,\beta=\frac{1}{T})$\,.}
\bea
\cS_{\rm bulk} &=& \frac{\beta \vol(\Sig_g)}{8\pi G_N} \int_{r_h}^{r_{\infty}} \frac{1}{2} \Blp e^{2(V-U)}(e^{2U})' \Brp' \label{ActionTotalDer}\,.
\eea
We now see that $\cS_{\rm bulk}$ combines nicely with the GHY term \eq{GHYterm}
\bea
\cS_{{\rm GHY}}&=& -\frac{\beta \vol(\Sig_g)}{8\pi G_N}\,e^{2V}(2V'-U'){\Big |}_{r=r_{\infty}}  \label{GHYterm2}
\eea
to give
\bea
\cS_{\rm bulk} +\cS_{\rm GHY} &=&  -\frac{A}{4 G_N}+ \frac{\beta \vol(\Sig_g)}{8\pi G_N} \Bslb -  e^{2U}(e^{2(V-U)})'   \Bsrb_{r=r_\infty}  \,,\label{actEvalbos}
\eea
where the area of the horizon is
\be
A=\vol(\Sig_g)  e^{2(V-U)} {\Big |}_{r=r_h}\,.
\ee
In deriving \eq{actEvalbos} we have used the definition of the inverse temperature
\be
\beta=\frac{4\pi}{(e^{2U})'}{\Big |}_{r=r_h}
\ee
and the simple but illuminating relation
\bea
e^{2V}(2V'-U') &=& e^{2U}(e^{2(V-U)})' +\frac{1}{2}e^{2(V-U)} (e^{2U})' \,.
\eea
The second term on the RHS of \eq{actEvalbos} is divergent and must be regularized, an additional subtle question is whether this term contributes a finite amount to the action. 

When evaluated on  a BPS solution using \eq{BPSW}, we find that \eq{actEvalbos} simplifies to
\bea
\cS_{\rm bulk} +\cS_{\rm GHY}  {\Big |}_{\rm BPS}&=&  -\frac{A}{4 G_N} - \frac{\beta \vol(\Sig_g)}{4\pi G_N} W {\Big |}_{r=r_\infty} \,, \label{SBPS}
\eea
where $W$ is the superpotential for the 1d action of section \ref{sec:BPSform}
\bea
W&=& e^{2V-U} \Im(e^{-i\psi} \cL )+ e^U \Re(e^{-i\psi} \cZ)\,. \label{superpot}
\eea
To prove our main result we would like to demonstrate that the second term on the RHS of \eq{SBPS} cancels against the required counterterms.

\subsection{BPS form of the action} \label{sec:BPSform}

A key step in our technique for computing the on-shell action is the BPS form of the action given in \cite{Dall'Agata:2010gj}. Starting with \eq{N2Bulk}, the authors of \cite{Dall'Agata:2010gj} dimensionally reduced this action to one dimension and recast it as a sum of squares plus boundary term\footnote{This 1d action must be supplemented by the {\it zero-energy} constraint and then the equations of motion will give \eq{eom1}-\eq{eom3} }:
\bea
\cS_{{\rm bulk}}&=& \cS_{\rm square} + \cS_{\rm bdy}  \label{firstorderaction}\\
&&\non\\
\cS_{\rm square} &=&  \frac{\beta \vol(\Sig_g)}{8\pi G_N } \int^{r_{\infty}}_{r_h} dr  
{\Big \{} -\frac{1}{2} e^{2(U-V)} \xi^T \cM \xi  -e^{2V} \bslb \psi'+\cA_r + 2e^{-U}  \Re (e^{-i\psi} \cL) \bsrb^2 \non \\
&&\non\\
&&- e^{2V}\bslb V' -2 e^{-U} \Im(e^{-i\al} \cL)\bsrb ^2 {\Big \}} \label{Ssquare} \\
 \cS_{\rm bdy} &=& \frac{\beta \vol(\Sig_g)}{8\pi G_N } \int^{r_{\infty}}_{r_h} dr  
{\Big \{} -(\kappa +\langle \cG,\cQ\rangle )-2\del_r W +\del_r \bslb e^{2V}(2V'-U') \bsrb{\Big \}} \label{Sbdy}
 \eea
where\footnote{We have corrected a small typo in \cite{Dall'Agata:2010gj} for the coefficient of $ (\psi'+\cA_r)$}
\bea
\xi &=& 2 e^{2V}   \Im\blp e^{-i\psi}e^{-U}\cV \brp '  + 4 e^{2V-U} (\psi'+\cA_r)  \Re \blp e^{-i\psi}\cV \brp + \cQ  + e^{2(V-U)} \Om \cM \cG \,.
\eea
The domain of integration is from the horizon $r_h$ where $e^{U}$ vanishes, to a large but finite asymptotic point $r_{\infty}$. The superpotential $W$ in \eq{Sbdy} is the same as in \eq{superpot}.

The Gibbons-Hawking-York \cite{Gibbons:1976ue, York:1972sj} boundary term takes the simple form \eq{GHYterm2} which is the same form and normalization as the second total derivative term in \eq{Sbdy} thus cancelling the contribution of this total derivative term at $r_{\infty}$. We also note that since $e^{V-U}$ and the scalar fields are finite at the horizon while $e^{U}$ vanishes there, we have
\be
W{\Big |}_{r=r_h}=0\,.
\ee
Finally we see that (with no reference to the BPS limit)
\bea
\cS_{\rm bulk}+ \cS_{\rm GHY}&=& \cS_{\rm square} -\frac{A}{4 G_N}  - \frac{\beta \vol(\Sig_g)}{4\pi G_N } W{\Big |}_{r=r_{\infty}}  \label{SBGHY}
\eea
and comparing with \eq{SBPS} we have 
\be
\cS_{\rm square}  {\Big |}_{\rm BPS}=0\,. \label{Ssqzero}
\ee

One might expect that \eq{Ssqzero} follows immediately from the fact that the {\it integrand} vanishes on BPS solutions, indeed this is the essence of the Bogolmony argument of \cite{Freedman:2013oja}, however this  intuition is sullied by the divergence coming from the integral over Euclidean time. Indeed, for the BPS universal black hole, while it is true that $\cS_{\rm square} =0$, it is the sum of two finite terms which cancel (which is possible because $\cM$ is negative definite). Explicitly, for the universal black hole we have the following contributions to $\cS_{\rm square}$ (the second term in \eq{Ssquare} vanishes exactly):
\bea 
\frac{\beta }{4\pi} \int^{r_{\infty}}_{r_h} dr \, \frac{1}{2} e^{2(U-V)} \xi^T \cM \xi &=&  - \frac{1}{96g^2} +\frac{\eta^{1/2}}{12\sqrt{2}\, g^{3/2}} +\cO(\eta)  \\
\frac{\beta }{4\pi} \int^{r_{\infty}}_{r_h} dr  \, e^{2V}\bslb V' -2 e^{-U} \Im(e^{-i\al} \cL)\bsrb ^2  &=&  \frac{1}{96g^2} +\frac{\eta^{1/2}}{48\sqrt{2}\, g^{3/2}} +\cO(\eta)
\eea
where we have used the relation for the universal black hole
\be
\frac{1}{T}=\frac{\beta}{4\pi}= \frac{1}{2^{7/2} g^{3/2}} \frac{1}{\eta^{1/2}} + \frac{1}{8g} + \cO(\eta^{1/2})\,.
\ee
So we see that in a perturbation around the BPS solution we have 
\be
\cS_{\rm square}= \cO(\eta^{1/2})
\ee
but the $\cO(\eta^0)$ term vanishes by a non-trivial cancellation.

\subsection{Cancellation of divergences}\label{sec:divergences}

We must add boundary counterterms to render the on-shell action finite. For minimal gauged supergravity it has been established some time ago \cite{Henningson:1998gx, Henningson:1998ey, Balasubramanian:1999re, Hyun:1998vg} that one should add 
\bea
\cS_{{\rm ct}}&=& \frac{1}{8\pi G_N} \int_{M_{\infty}}d^3x \sqrt{h}\, \Bslb \frac{1}{g}R_{(3)} + g  \Bsrb \,. \label{bdyctmin}
\eea
The generalization of \eq{bdyctmin} to include scalar fields has been studied \cite{Papadimitriou:2007sj} and quite recently revisited to include the constraints imposed by supersymmetry \cite{Freedman:2013oja, Freedman:2016yue}, following which we generalize the second term in \eq{bdyctmin} with part of the superpotential \eq{superpot}
\bea
\cS_{{\rm ct},\cL}&=& \frac{2}{8\pi G_N} \int_{M_{\infty}} d^3x \sqrt{h}\, \Im(e^{-i\psi}\cL)\,.\label{SctL}
\eea
canceling exactly the similar term in \eq{SBGHY}. 

The precise generalization of the first term in \eq{bdyctmin} is not immediately clear but it should be of the form 
\bea
\cS_{{\rm ct},R}&=& \frac{1}{8\pi G_N} \int_{M_{\infty}}d^3x \sqrt{h}\, Z(z^i)R_{(3)} \label{SctR}
\eea
where $Z(z^i)$ is a function of the scalar fields. 
To summarize, the action so far is given by 
\be
\cS_{\rm on-shell}=S_{{\rm bulk}}+\cS_{{\rm GHY}}+\cS_{{\rm ct},\cL}+\cS_{{\rm ct},R} \label{S1d}
\ee
and using 
\be
R_{(3)} =2\kappa e^{2U-2V}
\ee
we have the linearly divergent terms
\bea
\cS_{div}&\sim& \Bslb \kappa e^{U} Z(z^i) -\kappa r - \langle \cG,\cQ \rangle r -2  e^U \Re(e^{-i\al} \cZ) \Bsrb_{r=r_{\infty}}\,. \label{Sdiv}
\eea
It follows from \eq{BPS3} and section \ref{sec:AdS} that for black holes which asymptote to a supersymmetric AdS$_4$ vacuum, we have
\be
\Bslb\langle \cG,\cQ \rangle r+ 2  e^U \Re(e^{-i\al} \cZ) \Bsrb_{r=r_{\infty}} =0\,. \label{ReZidentity}
\ee
Since $e^{U}=\frac{r_{\infty}}{R_{\rm AdS}}+\cO(r_{\infty}^0)$, in order that we cancel the linear divergence we require
\be
Z(z^i) = R_{\rm AdS} + \cO(r_{\infty}^{-1})\,.
\ee
It may be desirable to find an expression for $Z$ in terms of the fields and not $R_{\rm AdS}$. It might be reasonable that $Z=\frac{1}{\Im (e^{-i\psi} \cL)}$, which would obey this asymptotic condition, is symplectic invariant and is local on field space but we do not see a way to precisely check this. Nonetheless, we will see below that for the evaluation of the on-shell action, the precise form of $Z(z^i)$ will not be needed beyond this leading order.

We note that the divergences in the four terms in \eq{Sdiv} are all equal in magnitude and we can consider that the first and second two terms cancel amongst themselves. As such we can relax the BPS Dirac quantization constraint \eq{DiracQu} to \eq{DiracQu2} while maintaining this cancellation; our analysis is valid also for the class of non-extremal black holes obtained by varying the charges in the solutions of \cite{Cacciatori:2009iz, Halmagyi:2014qza} to satisfy \eq{DiracQu2}.

We have shown a method to cancel the divergences for solutions which asymptote to a BPS AdS$_4$ background in the UV. It seems reasonable that the counterterms we add are universal in that all solutions to our theory should be regulated by the same set of counterterms. If so, it raises an issue of how to render finite a solution which in the UV does not satisfy the BPS conditions. Power counting suggests that $\cS_{\rm square}$ diverges as $\cO(r^3)$ for large $r$ and additional counterterms would be needed to cancel such a divergence.

\subsection{Finite action from the boundary}\label{sec:finitebdy}

Having cancelled the linear and cubic divergences from the second term on the RHS of \eq{SBGHY}, we are left with the sometimes thorny issue of $\cO(r_{\infty}^0)$ contributions to the action. For the class of BPS solutions we consider in this paper, there is a satisfying resolution of this: all such finite contributions cancel. 

The scalar fields $z^i$ have for large $r$ the expansion 
\be
z^i = z_0^i + \frac{z_1^i}{r}+ \frac{z_2^i}{r^2}+\ldots
\ee
and it follows that the expansion of $Z(z^i)$ and the central charge $\cZ(z^i,\cQ)$ are of a similar form
\bea
Z(z^i)&=& R_{\rm AdS} + \frac{Z_1(z^i_1)}{r_{\infty}}+ \frac{Z_2(z^i_1,z_2^i)}{r_{\infty}^2}+\ldots \\
\cZ(z^i,\cQ)&=&-\frac{1}{2}\langle \cG,\cQ \rangle  + \frac{\cZ_1(z^i_1,\cQ)}{r_{\infty}}+ \frac{\cZ_2(z^i_1,z_2^i,\cQ)}{r_{\infty}^2}+\ldots
\eea
where $(Z_1,\cZ_1)$ are homogeneous of degree one\footnote{so that $f-x\frac{\del f}{\del x}=0$} in $z_1^i$, $(Z_2,\cZ_2)$ are degree one in $z_2^i$ and degree two in $z_1^i$.
For a general black hole solution $e^U$ may in addition have a constant term in its expansion (although for the magnetic solutions of \cite{Cacciatori:2009iz} it does not) but this will not be important for our purposes.

The procedure for enforcing Neumann boundary conditions on the scalar fields is well known \cite{Klebanov:1999tb}: we should perform a Legendre transform
\bea
\cS_{\rm Legendre} &=& -\int d^3x \, z_1^i\frac{\delta  \cS_{\rm on-shell}}{ \delta z_1^i}\,,\\
\tcS_{\rm on-shell}&=& \cS_{\rm on-shell}+\cS_{\rm Legendre} \label{Legendre}
\eea
and it is $\tcS_{\rm on-shell}$ which is related to the boundary partition function. 

We see again the utility of the first order action \eq{firstorderaction}: $\cS_{\rm square}$ vanishes in the BPS limit while $\cS_{\rm bdy}$ contributes finitely to $\frac{\delta  \cS_{\rm on-shell}}{ \delta z_1^i}$. Since the superpotential term proportional to $\Im (e^{-i\psi }\cL)$ has been exactly cancelled, it does not contribute to lower order divergences. As such we have the crucial observation that $z_2^i$ does not contribute to the $\cO(r_{\infty}^0)$ part of $\cS_{\rm Legendre}$. The vanishing of $\cS_{\rm square}$ is central to this argument, even if we would have somehow cancelled its cubic and linear divergences, we would risk the presence of finite terms which depend non-linearly on the modes $z_1^i$ and indeed we expect this when the $\eta$-mode scales as $\frac{1}{r^3}$ at the boundary. More precisely we have
\bea
\cS_{\rm on-shell}{\Big |}_{\rm BPS}
&\sim & -\frac{A}{4 G_N} -\frac{ \beta \vol(\Sig_g) }{8 G_N} \Blp \frac{r_{\infty}}{R}\Bslb  \kappa \frac{Z_1(z_1^i)}{r_{\infty}} -2 \frac{\cZ_1(z_1^i,\cQ)}{r_{\infty}} \Bsrb  +\cO(r_{\infty}^{-1})\Brp
\eea
and since $Z_1$ and $\cZ_1$ are homogeneous of degree one in $z_1^i$, it follows that in the BPS limit the $\cO(r_{\infty}^0)$ term in $\cS_{\rm Legendre}$ cancels the $\cO(r_{\infty}^0)$ term in the asymptotic expansion of the $\cS_{\rm on-shell}$. 

This concludes the argument that the contribution to the Lagrangian from terms of order $\cO(r_{\infty}^0)$ vanishes for BPS solutions. For non-BPS solutions our argument breaks down in several interesting ways but assuming the non-BPS parameter scales as $\cO(r_{\infty}^{-3})$ at the boundary (such as the finite termperature perturbation) no new divergences are introduced. However we do expect a non-vanishing contribution of order $\cO(r_{\infty}^0)$ coming from $\cS_{\rm square}$. Such a contribution  will be non-linear in the $z_1^i$ modes and as such will not be cancelled by the Legendre transform.

\subsection{The complete on shell action}
We have now shown that the correct on-shell action is given by 
\be
\tcS_{\rm on-shell}=\cS_{\rm bulk}+ \cS_{\rm GHY} +\cS_{{\rm ct},\cL}+\cS_{{\rm ct},R}+ \cS_{\rm Legendre}
\ee
and that
\be\fbox{$
\tcS_{\rm on-shell} {\Big |}_{\rm BPS}=-\frac{A_h}{4 G_N}
$}\,. \label{mainresult}
\ee
We repeat the steps here for clarity
\begin{enumerate}
\item Evaluate the bulk action plus the GHY term on-shell using the second order equations of motion for the metric, giving \eq{actEvalbos}
\item Evaluate \eq{actEvalbos} on a BPS solution giving \eq{SBPS}
\item Reduce the four dimensional theory to a one dimensional action \eq{firstorderaction}
\item By comparison with \eq{SBPS}, show that terms in the one dimensional action which are sums of squares must cancel amongst themselves  on BPS solutions
\item Cancel cubic divergences in the action by adding part of the superpotential \eq{SctL} as a counterterm
\item Cancel linear divergences by adding the boundary Ricci scalar multiplied by a function of the scalar fields \eq{SctR}
\item Cancel  terms finite at the boundary in the on-shell Lagrangian in the BPS limit by performing the Legendre transformation \eq{Legendre} and enforcing Neumann boundary conditions on the scalars
\item The remaining term in the on-shell action comes from the horizon and is exactly the entropy of the black hole

\end{enumerate}

\section{Conclusions}

We have shown that in a large class of gauged supergravity theories, the on-shell action is equal to minus the entropy of the black hole \eq{mainresult}. Since our line of reasoning did not require the explicit solution beyond the AdS$_4$ boundary conditions, it should be straightforward to generalize these arguments to theories with charged hypermultiplets where solutions are harder to find \cite{Halmagyi:2013sla, Erbin:2014hsa, Chimento:2015rra, Klemm:2016wng, Monten:2016tpu, Guarino:2017eag} but there are more avenues for embedding such theories into string or M-theory. It would also seem likely that our methods will generalize to asymptotically AdS solutions in other dimensions.

Holography relates the on-shell action of the gravitational theory to the field theory partition function by a semi-classical approximation to the free-energy\footnote{We thank Ioannis Papadimitriou for discussions regarding \cite{Papadimitriou:2005ii}}
\bea
Z= e^{-\beta F} \sim e^{-\cS_{{\rm on-shell}}}\,.
\eea
Indeed, the Gibbs free energy of the gravity theory \cite{Gibbons:1976ue} in AdS spacetimes was shown to obey the quantum statistical relation \cite{Papadimitriou:2005ii}:
\be
\cS= \beta G(T,\Om_i,\Phi)\,,\qquad G\equiv M-TS-\Om_iJ_i-\Phi Q\,. \label{QSR}
\ee
The additional insight of holography is of course that $G$ is the free energy of a specific boundary quantum field theory. From \eq{QSR} it would seem that for the on-shell action to equal (minus) the entropy we should obey some kind of BPS-like bound 
\be
M-\Om_iJ_i-\Phi Q=0 \label{BPSbound}\,.
\ee
In this paper we have shown explicitly that for static-BPS black holes in AdS$_4$ within a certain class of gauged supergravity theories, the on-shell action is equivalent to minus the entropy. It would be interesting to relate this directly to a BPS bound like \eq{BPSbound}.

\vspace{1cm} \noindent {\bf Acknowledgements:} We would like to thank Monica Guica, Ioannis Papadimitriou, Boris Pioline and Ashoke Sen for discussions. This work was conducted within the ILP LABEX (ANR-10-LABX-63) supported by French state funds managed by the ANR within the Investissements d'Avenir program (ANR-11-IDEX-0004-02) and supported partly by the CEFIPRA grant 5204-4. The work of SL is supported by the Marie-Sklodowska
Curie Individual Fellowship 2014.

\begin{appendix}

\section{Special Geometry Background}\label{app:Special}

Here we summarize our conventions from special geometry, we use the same conventions as \cite{Halmagyi:2014qza} where more detailed information can be found.

The symplectic sections are given by
\be
\cV=\bpm L^\Lam \\ M_\Lam \epm \ =\ e^{K/2} \bpm X^\Lam \\ F_\Lam  \epm \label{cVdef}
\ee
where
\be
X^\Lam=\bpm1 \\ z^i \epm\,,\quad\quad z^i=x^i+iy^i
\ee
and satisfy
\be
\langle \cV,\cVbar \rangle = -i\,,\quad\quad \langle D_i\cV,D_{\jbar}\cVbar \rangle=ig_{i\jbar}
\ee
where the symplectic inner product is 
\be
\langle A,B\rangle= B^\Lam A_\Lam-B_\Lam A^\Lam.
\ee
Any symplectic vector can be expanded in these sections, for example the charges are expanded as
\be
\cQ= i \cZbar \cV - i \cZ \cVbar + i \cZbar^{\ibar} D_{\ibar} \cVbar - i \cZbar^i D_i\cV
\ee
where 
\be
\cZ=\langle \cQ,\cV\rangle\,,\quad\quad \cZ_i = \langle \cQ,D_i \cV \rangle\,.
\ee
The other symplectic invariants which we use are constructed from the gauge couplings
\bea
\cL=\langle \cG,\cV\rangle\,,\quad\quad \cL_i = \langle \cG,D_i \cV \rangle\,.
\eea
We also have a complex structure on the symplectic bundle over $\cM_v$:
\be
\Om\cM \cV= -i \cV\,,\quad\quad \Om\cM (D_i \cV)= i D_i \cV
\ee
where 
\be
\Om=\bpm0 & -1\!\!1 \\ 1\!\! 1 & 0 \epm\,,\quad\quad\cM=\bpm 1 & -\cR \\ 0 & 1 \epm\bpm \cI & 0 \\ 0 & \cI^{-1} \epm\bpm 1 & 0 \\ -\cR & 1 \epm
\ee
and $\cN=\cR+i\cI$ is the standard matrix which gives the kinetic and topological terms in the action for the gauge fields.

\end{appendix}

\providecommand{\href}[2]{#2}\begingroup\raggedright\endgroup
\end{document}